\documentclass[epj]{svjour}
\usepackage{graphics}
\usepackage{amsfonts}
\usepackage{subfigure}
\usepackage{pslatex}
\usepackage{ae,aecompl}
\usepackage{epsfig}
\usepackage{graphics}
\usepackage{graphicx}
\usepackage{amsmath}
\usepackage{amssymb}
\usepackage{pifont}
\usepackage{psfrag}
\usepackage{subfigure}
\usepackage{float}
\usepackage{color}

\begin{document}


\title{Segmentation algorithm for non-stationary compound Poisson processes}
\subtitle{With an application to inventory time series of market members in a financial market}
\author{Bence T\'oth\inst{1}\thanks{E-mail: bence@santafe.edu} \and Fabrizio Lillo\inst{1,2} \and J. Doyne Farmer\inst{1,3}
}                     
%
%
\institute{Santa Fe Institute, 1399 Hyde Park Road, Santa Fe, NM 87501, USA \and
Dipartimento di Fisica e Tecnologie Relative, Universit\'a di Palermo, Palermo, Italy \and
LUISS Guido Carli, Viale Pola 12, 00198, Roma, Italy}
\date{Received: date / Revised version: date}
%
\abstract{We introduce an algorithm for the segmentation of a class of regime switching processes. 
The segmentation algorithm is a non parametric statistical method able to identify the regimes (patches) of a time series.
The process is composed of consecutive patches of variable length. In each patch the process is described by a stationary compound Poisson process, i.e. a Poisson process where each count is associated with a fluctuating signal. The parameters of the process are different in each patch and therefore the time series is non-stationary.  Our method is a generalization of  the algorithm introduced by Bernaola-Galv\'an, {\it et al.}, Phys. Rev. Lett., \textbf{87}, 168105 (2001). We show that the new algorithm outperforms the original one for regime switching models of compound Poisson processes. As an application we use the algorithm to segment the time series of the inventory of market members of the London Stock Exchange and we observe that our method finds almost three times more patches than the original one. 
} 
\maketitle




\advance\baselineskip by 3pt

\section{Introduction}
\label{intro}

Many time series from natural and social phenomena exhibit non-stationarity. A proper detection and characterization of this non-stationarity is a major challenge in time series analysis. Among the possible types of non-stationarities, regime switching (or mosaic organization) plays a major role. In regime switching models the parameters of the model change abruptly from time to time and the time series is organized in consecutive patches, each characterized by a distinct set of model parameters. There is a vast literature on regime switching models \cite{hamilton} perhaps the best known of which are Hidden Markov Models \cite{rabiner}. Most models describe processes in discrete time, although there are also models in continuous time (i.e point processes) \cite{chib1998}. 

Regime switching models have been applied to a large variety of systems. One of the first applications comes from research in quality control, where one wishes to detect deviations from an expected output level of a production process by observing a signal \cite{shewhart31}.  A more recent example is human heartbeat interval dynamics. Studies showed \cite{bernaola-galvan2001} that the time series of heartbeat intervals are organized in consecutive temporal segments with different local mean heart rates. Proper segmentation of the heartbeat data  can bring up relevant physiological information, with the parameters differentiating between healthy and ill patients.  Other examples include changes in economic regression models; financial data analysis; transport networks on which the flow of some quantity can be studied (electric-energy networks, internet traffic); geological data and seismic signal processing (appearance of tsunamis and earthquakes); epidemiology; statistical image processing; and the appearance of shock wave fronts \cite{carlstein}.

An example that we will consider in this paper is the inventory time series of financial market members. Market members trading a large amount of shares tend to split their order into smaller transactions in order to limit
their own impact on the market \cite{Lillo05b,vaglica2008,toth}. This strategic order splitting leads to long regimes when a market member is consistently buying or selling. By studying the inventory time series (the number of stocks owned at any given moment by a market member) these regimes can be identified and important information on traders' behavior can be assessed.

Given an empirical time series assumed to be described by a regime switching process, the model fitting must be able to determine both the boundaries between consecutive patches and the model parameters in each patch. Given the probabilistic nature of the model, the boundary between two consecutive regimes is not directly observable and must be inferred from the data.  A segmentation algorithm is a statistical method able to identify the different regimes (patches) of a regime switching time series.

Here we generalize an existing algorithm \cite{bernaola-galvan2001}, originally developed to identify non-stationarity in the heart rate, as a method for segmenting regime switching processes where each regime can be characterized by a compound Poisson process. A compound Poisson process is a point process characterized by a rate (like a Poisson process) and a signal intensity distribution. In a normal Poisson process the signals are always of unit intensity. In a compound Poisson process the associated counting process (i.e. neglecting the intensity of the signals) is a Poisson process and the signal intensities are independent and identically distributed random variables and are also independent of the counting process. A compound Poisson process is stationary and is the simplest example of a so-called marked point process\footnote{A marked point process is a point process where both the counting process (rate) and the signal intensity (jump size) process are generic stochastic processes. Moreover, in general, in a marked point process  the counting process and the jump process are not independent \cite{Last}.}.  Physical examples include the celebrated Continuous Time Random Walk \cite{montroll}, which is the integral of a marked point process. Marked point processes have been applied to many different fields, ranging from earthquakes \cite{holden} to financial time series \cite{englerussel,scalas,montero}. 

In this paper we consider non stationary regime switching stochastic processes in which in each patch the time series is described by a compound Poisson process. The parameters of the process are different in different patches. The length of the patches where the process is stationary is a random variable with a given distribution. Moreover the length of the patches is described by an independent and identically distributed stochastic process. Note that the length of the patches is not necessarily exponentially distributed because the Poisson nature of the process is inside each patch and it is not describing the boundaries between different patches. 

Regime switching models of compound Poisson process have several different applications. One case is the disorder problem formulated by Kolmogorov \cite{kolmo}. In a disorder problem one has to detect as quickly as possible a change in the probabilistic properties of the observed process. Natural applications of this problem are quality control or any case in which an alarm has to be raised quickly. The compound Poisson disorder problem has been widely studied in the literature  \cite{galchuk71,peskir00,gapeev05,dayanik06}. In these processes  either the arrival rate, or the jump distribution, or both changes abruptly at an unknown and unobservable time. The type of studied processes is the same as the one investigated in our paper. The difference is that, while in the disorder problem one investigates the process in real time and tries to detect a regime shift quickly, in our case we consider the whole time series and find the different regimes ex-post. Other typical applications of compound Poisson processes and regime shifts are earthquakes \cite{aktas}, meteorological data \cite{palmer}, and packets in Ethernet traffic \cite{taqqu}. All these cases can be described by compound Poisson processes, and the identification of regime shifts helps in identifying systemic changes in the generating process. 

The segmentation algorithm that we introduce here  is a generalization of the method introduced in \cite{bernaola-galvan2001}. This is a top-down method, i.e. it first splits the whole time series in two subsets and then continues iteratively by breaking the series down to a more and more refined partition.  
One of the advantages of this method and of  our generalization is that they are non parametric methods, i.e. they do not postulate a known distribution of the patch length and of the signal intensity (or jump size). The method of Ref. \cite{bernaola-galvan2001} has been recently used in Refs. \cite{vaglica2008,moro2009} for segmenting financial time series of the market members' trading activity in the Spanish Stock Exchange and in the London Stock Exchange (LSE). 
In this paper we will also consider the application of our algorithm to the segmentation of inventory time series of LSE market members. In order to do this in a financially reasonable way, we generalize the null model considering the possibility of long inactivity periods between two consecutive patches.

The paper is organized as follows. In Section \ref{null-model} we discuss our null-models of time series. In Section
\ref{method} we introduce our segmentation methods and present simulation results.
We present empirical results for the inventory of market members at the LSE in Section \ref{results} and finish the paper with conclusions in Section \ref{conclusions}.

\section{Null models}
\label{null-model}

We present two variants of the null model of time series that we expect to segment. The first is a pure regime switching model of compound Poisson processes, while in the second there may be long inactivity periods between consecutive active patches. This second model is designed to better describe the financial time series investigated in Section \ref{results}.

In the first model the patches with different Poisson rate of activity follow each other.
We consider the simplest case in which the jump size can be $\pm 1$. Since jump sizes are independently and identically distributed random variables, the jump sizes in a patch are statistically characterized by one probability (say, the probability of a jump size equal to $+1$). The sign associated with the largest probability is  the dominant sign of the patch.

In the following we present results where the length of the patches, $T_{act}$, is distributed according to a lognormal distribution.
\begin{equation}
T_{act}=C_{act}e^{N(1,1)},
\end{equation}
where $C_{act}$ determines the characteristic length of the patches, and $N(1,1)$ is a normal distribution with mean $\mu=1$ and $\sigma=1$. The choice of a lognormal distribution is made in order to have a fat tailed distribution of regime lengths as in many real world systems. However the choice of a different distribution for the patch lengths (e.g. exponential distribution) does not affect the quality of the results.
The mean and standard deviation of $T_{act}$ are
\begin{eqnarray}
<T_{act}>=e^{3/2}~ C_{act} \simeq 4.48 ~C_{act}\\
\Delta T_{act}= \sqrt{(e^1-1) e^3} ~C_{act} \simeq 5.87 ~C_{act}
\end{eqnarray}
In order to generate a realization of the process, for each patch we choose a patch length $T_{act}$, a dominant sign ($\pm 1$ randomly), a noise level $0\leq\eta\leq 1$, and a rate $\alpha$. The dominant sign and the noise level determine the jump process. With probability $1-\eta$ the jump has size equal to the dominant sign and with probability $\eta$ it has the opposite sign.  The rate $\alpha$ determines the counting process, i.e. the probability per unit time that an event occurs is $\alpha$. This parameter is drawn in each patch from a uniform distribution in the interval $[0.5-\delta, 0.5+\delta]$, where $0\leq\delta\leq0.5$ is the dispersion of the rate. For each patch we generate a compound Poisson time series of length  $T_{act}$ with the chosen parameters.

\begin{figure}
\begin{center}
\includegraphics[angle=0,width=0.450\textwidth]{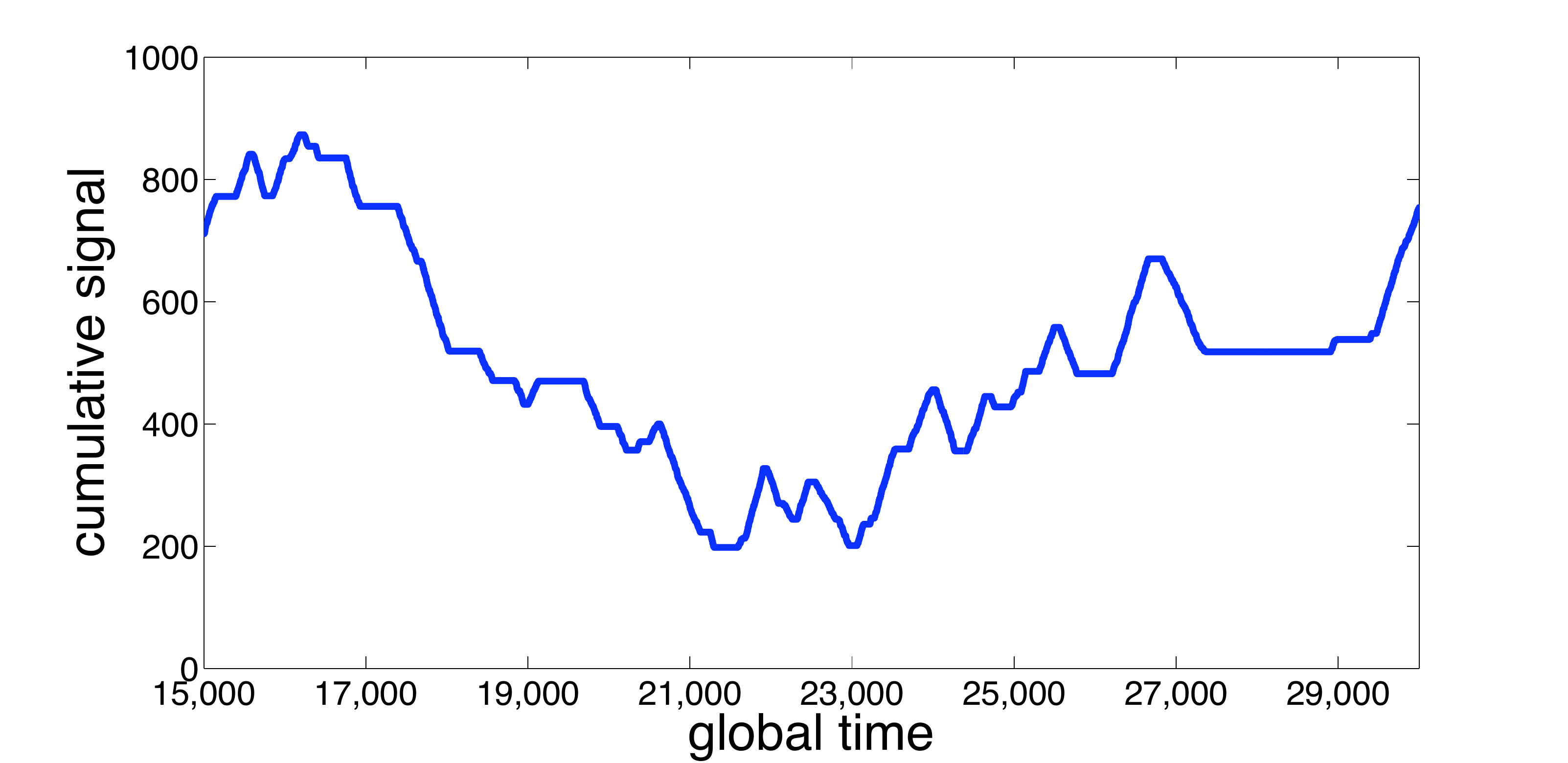}
\caption{\label{fig1} Snapshot of the integral of a simulated time series of a regime switching models of compound Poisson processes with inactivity periods, with parameters $C_{act}=50$, $C_{inact}=30$, $\delta=0.2$ and $\eta=0$.}
\end{center}
\end{figure}

In the second type of null model we also add inactive patches, i.e. patches when no jumps are present between each pair of consecutive active patches.  The active patches are generated in the same manner as before.
Inactive patches have length $T_{inact}$ distributed according to a lognormal distribution: 
\begin{equation}
T_{inact}=C_{inact}e^{N(1,1)},
\end{equation}
where, as before, $C_{inact}$ determines the characteristic length of the patches. Again, choosing some other distribution for the patch lengths does not affect the results.
Figure \ref{fig1} shows a snapshot of the integral of the time series.
In our simulations we varied the ratio $<T_{inact}>/<T_{act}>=C_{inact}/C_{act}$ between the mean length of active and inactive patches, sweeping the ratio from $0.02$ (almost
negligible length of inactivity) to 1 (equal mean length of active and inactive patches). 

We simulate the process in discrete time with steps of unit time length. Moreover we consider two different times for the process. The first is the \emph{global time}  and is the (coarse grained) real time. The second is the \emph{local time} in which we consider only the times when an event occurs. In other words, in local time we discard all the times when there are no events and end up with a time series containing only $\pm 1$. Local time discards all the information related to the point process nature of the time series and preserves only the jump (event) process. 

\section{Segmentation method}
\label{method}

\subsection{Segmenting regime switching models of compound Poisson processes}

As we have mentioned above, our segmentation algorithm is based on the algorithm in Ref. \cite{bernaola-galvan2001}. This algorithm works as follows. We move a sliding pointer over the series and for each position of the pointer we measure the mean of signal in the subset to the left and to the right of the pointer. We compute the statistic 
\begin{equation}
\label{eq5}
t=|(\mu_{left}-\mu_{right})/s_{D}|,
\end{equation}
where 
\begin{equation}
s_D=[(s_{left}^2+s_{right}^2)/(N_{left}+N_{right}-2)]^{1/2}(1/N_{left}=1/N_{right})^{1/2}
\end{equation}
is the pooled variance \cite{numrec}, $\mu_{left}$ and $\mu_{right}$ are the mean of the signal,
$s_{left}$ and $s_{right}$ the standard deviations and $N_{left}$ and $N_{right}$ the number of data points to the left and right of the pointer, respectively. We search for the position of the pointer for which the $t$ statistic of Eq. \ref{eq5} is maximal ($t_{max}$).   We make a cut if the significance of $t_{max}$ exceeds a given threshold, which we set to 99\% as in Ref. \cite{vaglica2008}. Note that one has to modify the relation between the statistic and the p-value according to  \cite{bernaola-galvan2001}.    After the cut we continue the method recursively on the newly created subsequences, until no further cut can be made. It is important to note that before a new cut is accepted we compute the modified $t$ value between the new segments and their neighbors and check if both values exceed the above significance level. 

One direct method for segmenting a compound Poisson process is to consider the time series in local time and run the algorithm of Ref. \cite{bernaola-galvan2001}. We will refer to this method as the \textit{local time t-test} throughout the paper. This approach has been used in Ref. \cite{vaglica2008} for segmenting inventory time series in financial markets (see also Section \ref{results}). However this approach is unable to identify changes in rate due to a regime shift because all the information on the rate is lost when one considers the compound Poisson process in local time.

We introduce a new method for segmenting compound Poisson processes, which we call the \textit{global time t-test}.
The idea is to apply the algorithm to the time series in global time, i.e. our series will be composed by many zeroes (in the coarse grained time intervals, when no event occurs) and $\pm 1$ (when one event occurs).
Naturally the power of the segmentation method depends considerably on the noise level $\eta$ and on the variance of the rate of Poisson processes inside different patches.
In order to study this we made simulations for different values of the dispersion of the rate of the Poisson process in different patches. As specified in the previous section, the rate is drawn from a uniform distribution in the interval $[0.5-\delta, 0.5+\delta]$ and we vary $\delta$. 
After generating the time series we run the local time t-test and the global time t-test. We compare the segmentation made by each method with the true segmentation.  All the results are the average of $100$ simulation runs and each run is made of $100$ patches.

For the assessment of the segmentation methods  we compute the Jaccard index \cite{jaccard1901} between the detected segmentation and the true segmentation. 
We denote the number of point pairs that are in the same patch both in the true time series and the segmented series by $M_{11}$. The number of point pairs that are in the same patch in the true time series but not in the segmented series is $M_{10}$, and similarly the number of pairs in different patches in the true time series but in the same patch in the segmented series is $M_{01}$. The Jaccard index is defined as
\begin{equation}
J=\frac{M_{11}}{M_{10}+M_{01}+M_{11}}.
\end{equation}
A perfect segmentation has $J=1$. 
In our case there are two possible ways of computing the Jaccard index. The first is to compute the Jaccard index in local time, i.e. by restricting the comparison to the time series where we have removed all the zeroes. We denote this measure by $J_{Local}$. The second way takes into account all time steps (even when there are no counts) and we denote it by $J_{Global}$. $J_{Local}$  tells us what fraction of the signals are segmented well, while $J_{Global}$ tells us, what fraction of global time is segmented well.

\begin{figure}
\begin{center}
\includegraphics[angle=0,width=0.450\textwidth]{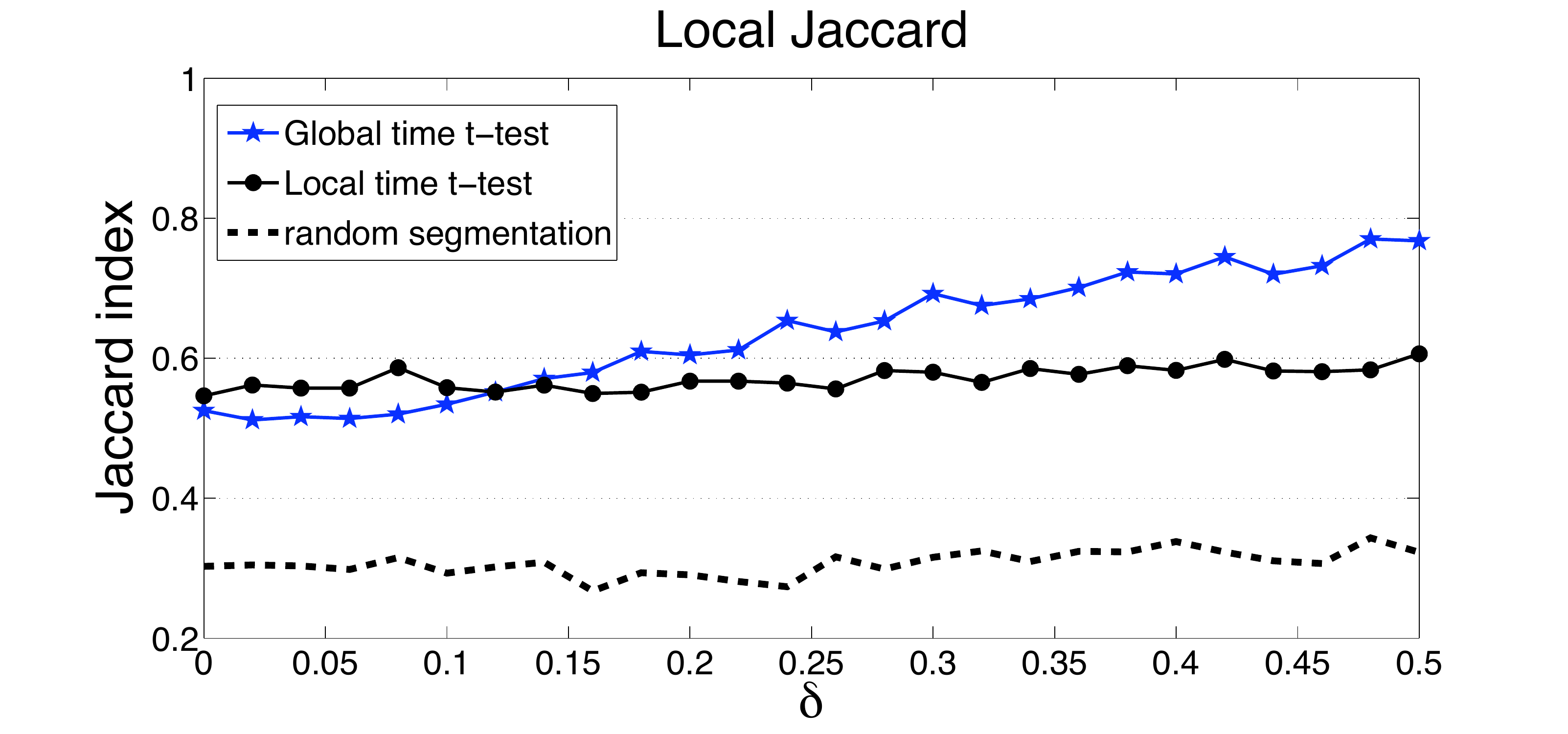}
\caption{\label{fig0.1} (Color online) The local Jaccard index $J_{Local}$ as a function of the dispersion $\delta$ of the rate $\alpha$ 
for the local time t-test (black circles), and the global time t-test (blue stars). 
The black dashed line shows the Jaccard index for random segmentation. Here $\eta=0$.}
\end{center}
\end{figure}

\begin{figure}
\begin{center}
\includegraphics[angle=0,width=0.450\textwidth]{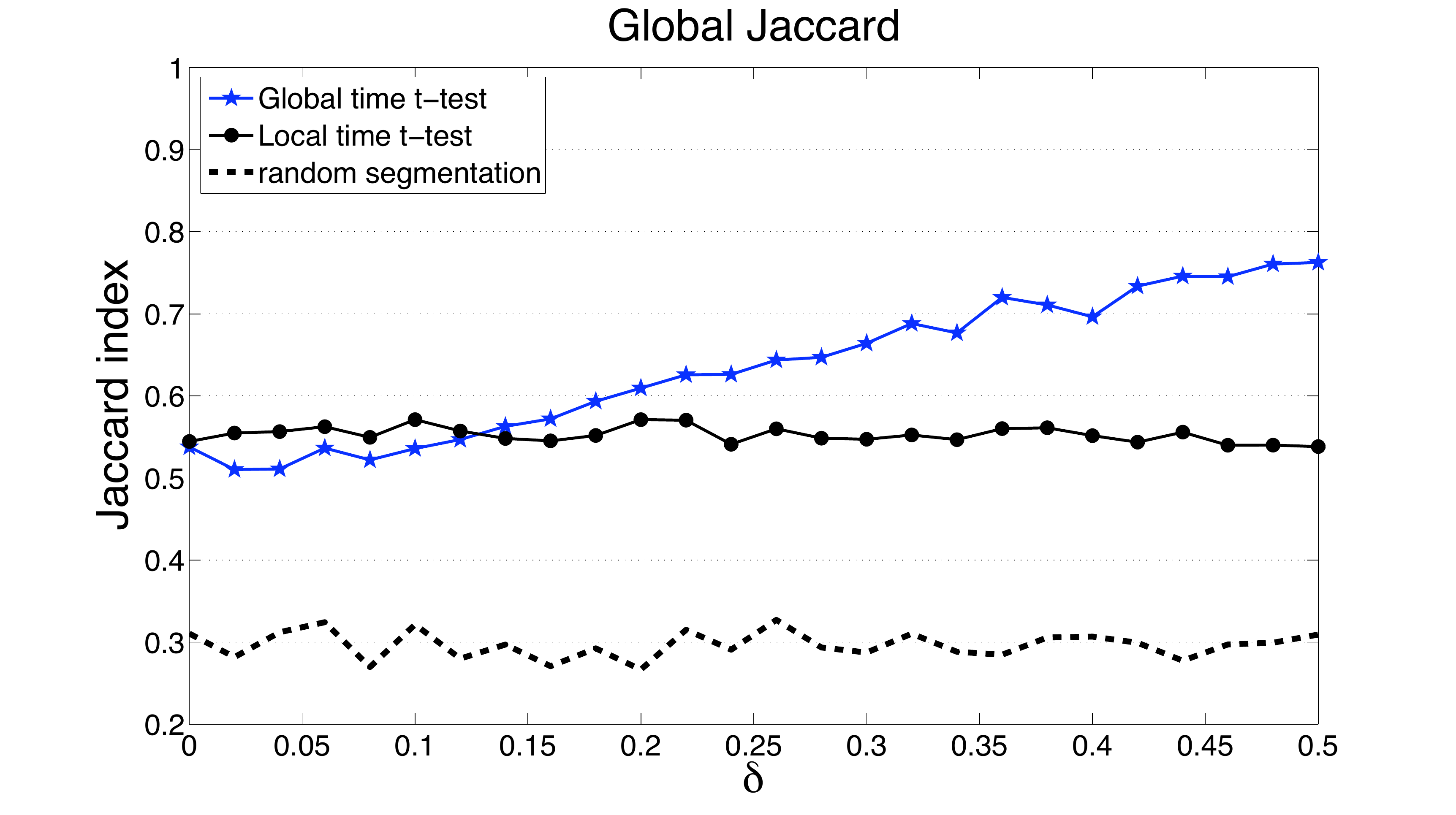}
\caption{\label{fig0.2} (Color online) The global Jaccard index $J_{Global}$ as a function of the dispersion $\delta$ of the rate $\alpha$ 
for the local time t-test (black circles), and the global time t-test (blue stars). 
The black dashed line shows the Jaccard index for random segmentation. Here $\eta=0$.}
\end{center}
\end{figure}

We studied both $J_{Local}$ and $J_{Global}$ for growing dispersion of the Poisson rates and different noise levels.
Figure \ref{fig0.1} shows $J_{Local}$ for the local time t-test  and the global time t-test. We also included the results that come from entirely random segmentation of the time series, applying the same number of cuts as the local time t-test. In this figure we set $\eta=0$. We can clearly see that for low values of $\delta$ the two methods perform roughly the same, while with growing $\delta$ the global time t-test outperforms the local time t-test by a large margin. The same result is seen in Figure \ref{fig0.2} where we show $J_{Global}$.

\begin{figure}
\begin{center}
\includegraphics[angle=0,width=0.450\textwidth,height=135pt]{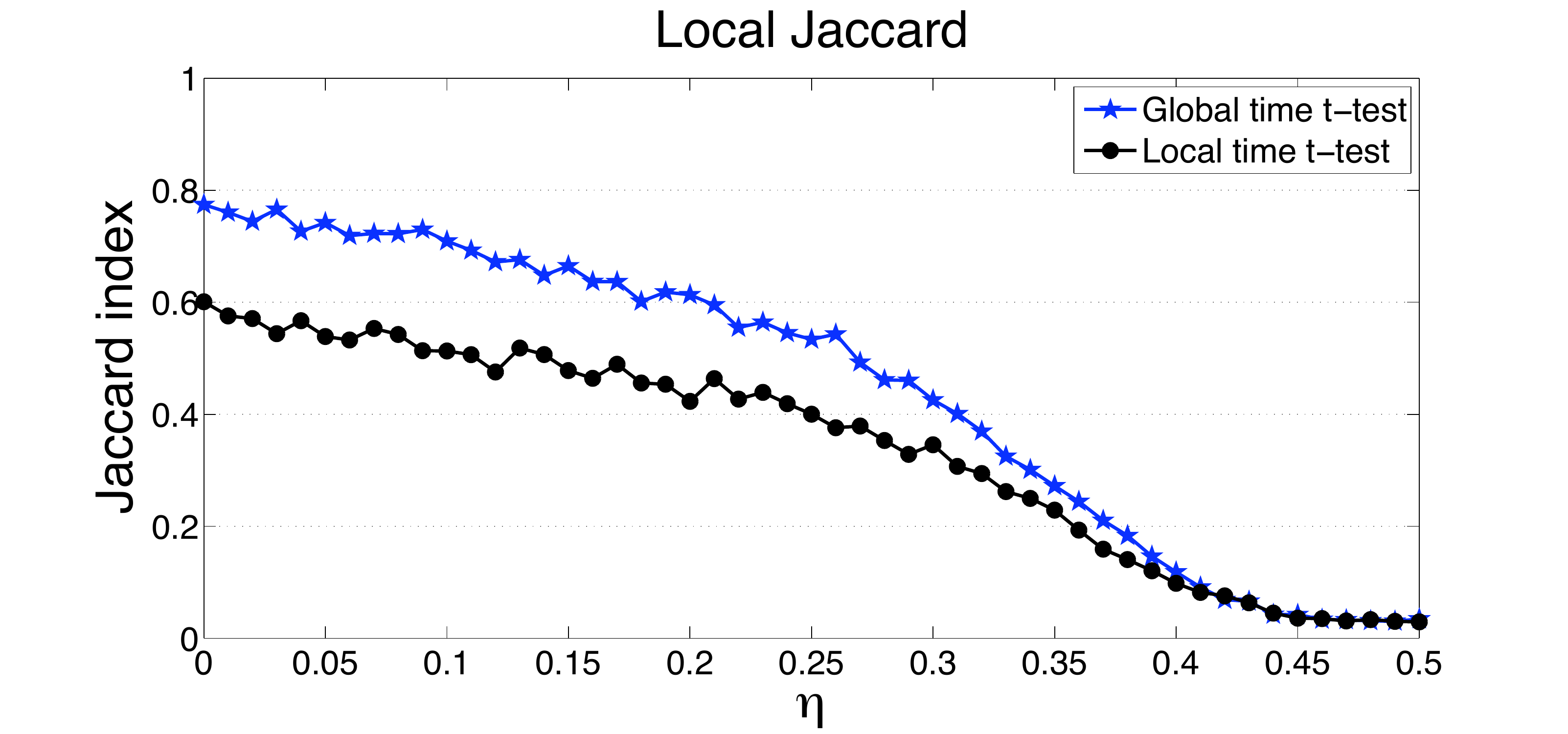}
\caption{\label{fig0.3} (Color online) The local Jaccard index $J_{Local}$ as a function of the noise $\eta$  
for the local time t-test (black circles), and the global time t-test (blue stars). Here $\delta=0.5$.}
\end{center}
\end{figure}

\begin{figure}
\begin{center}
\includegraphics[angle=0,width=0.450\textwidth,height=135pt]{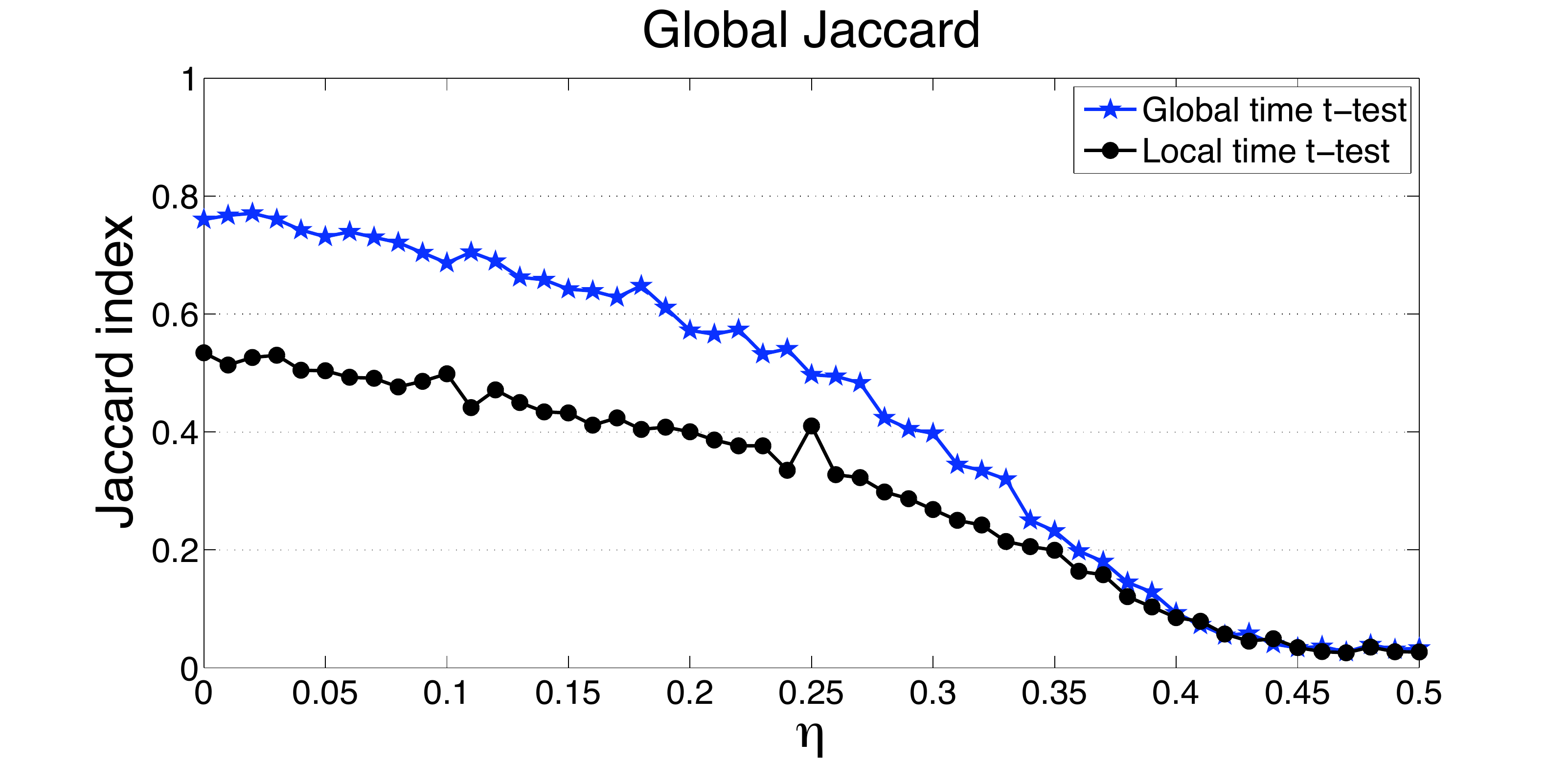}
\caption{\label{fig0.4} (Color online) The global Jaccard index $J_{Global}$ as a function of the noise $\eta$  
for the local time t-test (black circles), and the global time t-test (blue stars). Here $\delta=0.5$.}
\end{center}
\end{figure}

Figure \ref{fig0.3} shows the local Jaccard index  $J_{Local}$ for a model with $\delta=0.5$ and for different values of $\eta$. Clearly, $J_{Local}$ decreases as the noise level increases, but for all values of $\eta\lesssim0.4$ the global time t-test outperforms the local time t-test segmentation algorithm, and for $0.4\lesssim\eta\leq 0.5$ the two methods perform the same. A similar result is observed also with the global Jaccard index, as can be seen in Figure \ref{fig0.4}.

\subsection{Segmenting regime switching models of compound Poisson processes with inactivity periods}
\label{method2}

For regime switching models of compound Poisson processes with inactivity periods we modify the above algorithm. 
The segmentation algorithm, called the \textit{composite test}, has two modules.  First we run the global time t-test as described above. After the global time t-test, there is a second module that we call the \textit{rate test}: we check if the patches found by the global time t-test are consistent with the Poisson assumption. For each patch we estimate $\alpha$ as the inverse of the average inter-event time in the patch.
Then for each patch, we search for the longest period of inactivity between two events and we test
whether it is consistent with Poisson waiting times. Given a Poisson process with rate $\alpha$, the probability  of having at least one waiting time longer than $W$ is

\begin{equation}
\mathbb{P}(\omega_{max}>W)=1-(1-e^{-\alpha W})^{N},
\label{eq1}
\end{equation}
where $N$ is the number of jumps. 
Given the longest observed waiting time $\omega_{obs}$, we make additional cuts if the probability of having a waiting time $\omega_{max}\geq\omega_{obs}$ is less than a given threshold $q$  i.e.,

\begin{equation}
\omega_{obs}>\frac{\log(1-(1-q)^{1/N})}{-\alpha}.
\label{eq2}
\end{equation}
In this paper we use $q=0.01$. 
If Eq. \ref{eq2} holds, we make two additional cuts, one at the beginning and one at the end of the inactivity period and then we continue the process recursively on the new sequences.

To present the power of our composite test we show how it works for simulated time series of regime switching models of compound Poisson processes and inactivity periods as described in the previous section.
For activity periods we draw the Poisson rate randomly from the interval $\alpha\in [1/15, 1/5]$. We consider here the noiseless case ($\eta=0$), but we observe similar results for all values of $\eta$. After generating the time series we test the local time t-test segmentation method and the two methods introduced here, the global time t-test and the composite test. We compare the segmentations obtained by each method with the generated segments by computing the Jaccard index in local time and in global time.  As before, all results are the average of $100$ simulation runs.

\begin{figure}
\begin{center}
\includegraphics[angle=0,width=0.450\textwidth]{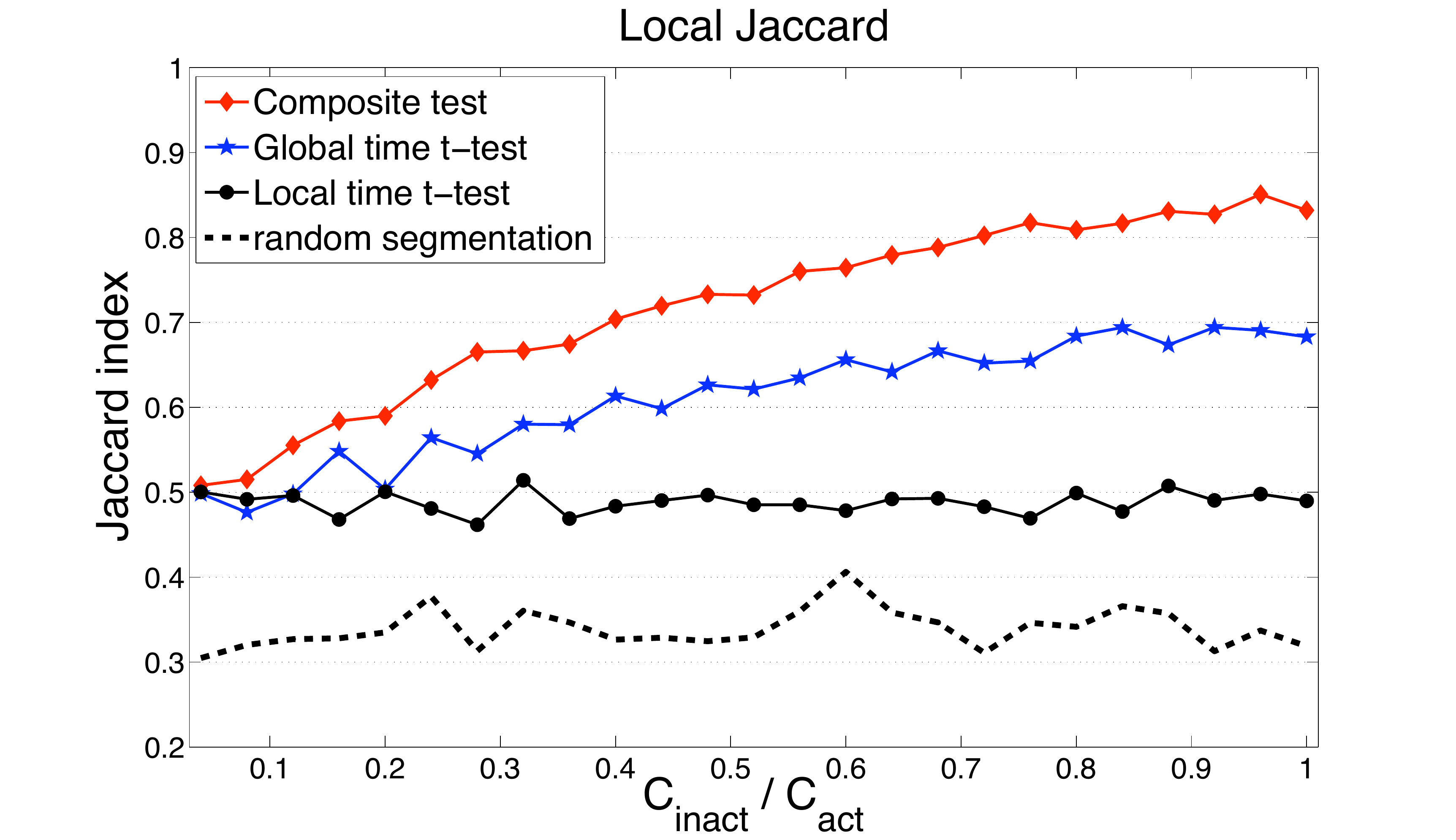}
\caption{\label{fig2} (Color online) The local Jaccard index $J_{Local}$ for the local time t-test (black circles), the global time t-test (blue stars) and the composite test (red diamonds). 
The black dashed line shows the Jaccard index for random segmentation.}
\end{center}
\end{figure}

Figure \ref{fig2} shows the results for $J_{Local}$. We present the $3$ methods, the local time t-test  (black circles), the global time t-test (blue stars) and the composite test (red diamonds). 
With the dashed line we also show the Jaccard index obtained by an entirely random segmentation,
applying the same number of cuts as in the local time t-test.
It can be clearly seen that the curve of the composite test is well above the other curves. For very short inactivity periods the $3$ methods perform roughly the same, while for growing length of inactivity periods the performance of the composite method (red diamonds) quickly becomes better.
The local time t-test (black circles) performs uniformly when changing the ratio of active and inactive patch length.

\begin{figure}
\begin{center}
\includegraphics[angle=0,width=0.450\textwidth]{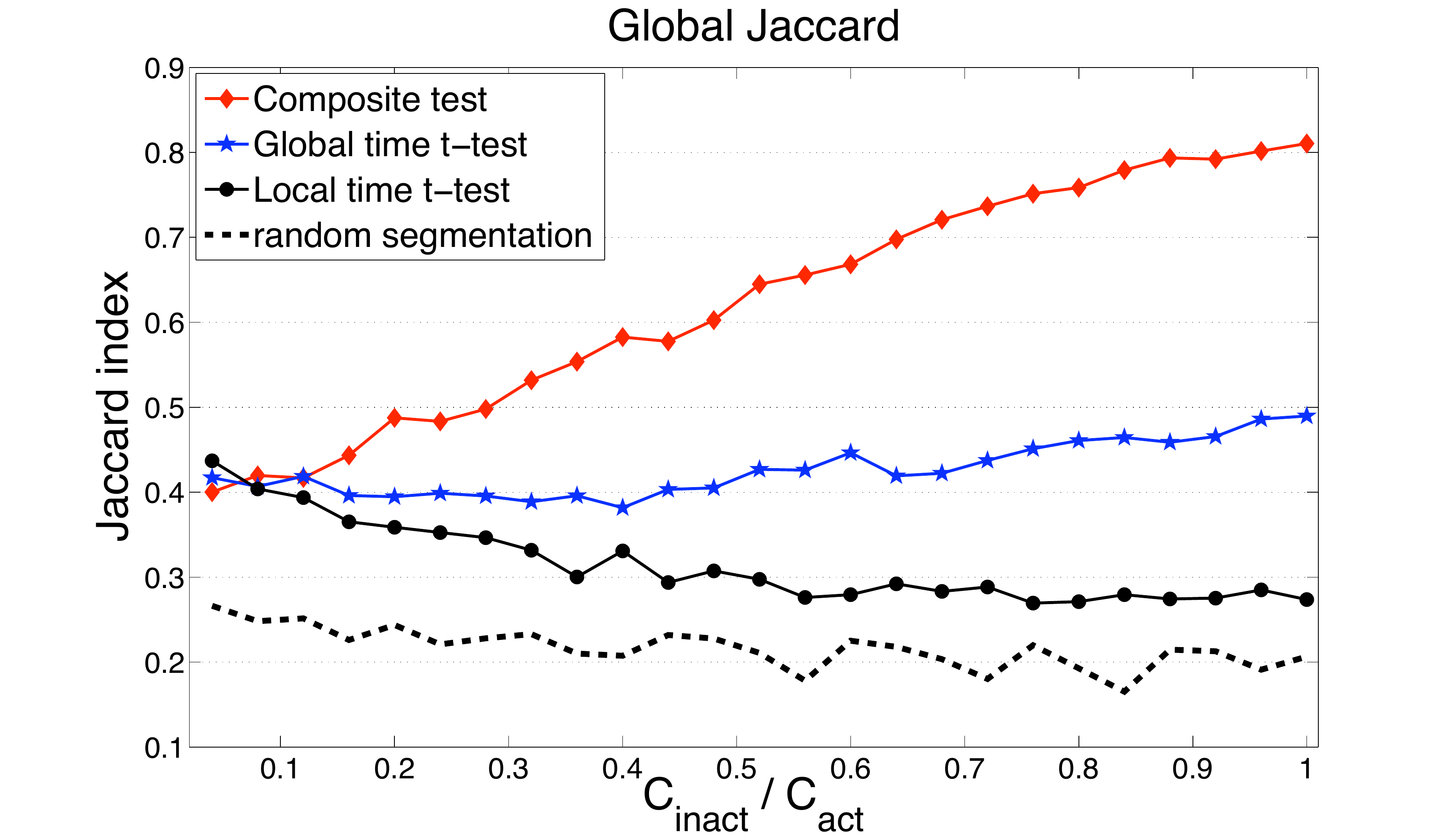}
\caption{\label{fig3}  (Color online) The global Jaccard index $J_{Global}$ for the local time t-test (black circles), the global time t-test (blue stars) and the composite test (red diamonds). 
The black dashed line shows the Jaccard index for random segmentation.}
\end{center}
\end{figure}

Figure \ref{fig3}  shows the results for $J_{Global}$. Again we see that for very short inactivity periods the $3$ methods perform roughly the same, while for longer inactivity patches the composite method (red diamonds) significantly outperforms the other methods. The global time t-test performs roughly the same as $C_{inact}/C_{act}$ changes, while the performance of the local time t-test declines as the relative length of inactivity periods increases.

From both plots we conclude that our proposed composite test outperforms the direct application of the original test to time series which can be described by regime switching models of compound Poisson processes (with or without inactivity periods). As already mentioned, the relative performance of the segmentation methods does not depend on the choice of the distribution of the patch lengths, $T_{act}$ and $T_{inact}$. The qualitative results do not change when using other distributions. This in fact is one of the strengths of our method: being a non parametric method, it does not postulate a known distribution of the patch length and of the signal intensity.

\section{Empirical analysis of financial data}
\label{results}

As already mentioned in the introduction, a real example where the proposed segmentation methods can be very useful is the analysis of financial inventory time series and the determination of hidden orders.

It has recently been shown that the signed market order flow, i.e. the time series of the sign of the market orders initiating trades, is a long memory process \cite{Lillo04,Bouchaud04}. It has been proposed that the long memory property is due to the practice of order splitting \cite{Lillo05b}. When a large investor decides to trade a large volume, it is quite unlikely that she places one large order in the market. What typically happens is that large trading orders are split into pieces and executed incrementally. We call these large orders {\it hidden orders}.   For strategic reasons traders attempt to keep the true size of their orders secret in order to minimize transaction cost and trade the order at a more favorable price (for a review of this problem, see \cite{bouchaud2009}).  Empirical evidence of a widespread practice of order splitting has been given in the empirical results of Refs \cite{chan1995,almgren2005,gallagher2006,vaglica2008,moro2009}.
More recently it has been shown more directly that order splitting is the main cause of the long memory of order flow \cite{toth}.

From an empirical point of view the detection of hidden orders is difficult because market participants do not reveal their trading intentions and strategy. It has been recently proposed to detect hidden orders from the analysis of the time series of the inventory of market members \cite{vaglica2008}. Inventory variation is the signed transaction volume of a given market member, where conventionally a buy (sell) transaction has a positive (negative) volume equal in absolute value to the number of shares traded.
Beside trading for themselves, market members act often as brokers or dealers. Therefore they are often not the real initiators of hidden orders, but the assumption of this approach is that it is unlikely that more than one large hidden order is given simultaneously to one broker. Under this assumption a large hidden order should be visible in the inventory time series of the broker as a random walk with a very strong drift. 

This approach has proven useful and has been applied to the determination of the market impact of hidden orders \cite{moro2009}. The segmentation method used in \cite{vaglica2008,moro2009} is the local time t-test of \cite{bernaola-galvan2001} applied to the time series of inventory variation of each broker {\it in local time}.  Now, if two consecutive buy hidden orders with the same typical volume per trade (largely determined by the liquidity present in the market)  are placed by the same broker with a time difference between the end of the first and the beginning of the second that is short compared to the length of the two hidden orders, a local time t-test segmentation method will identify them as a single hidden order, because the algorithm has no way of separating them. For this reason, the main claim of this paper is that a global time t-test segmentation method should work better in identifying hidden orders. 

One possible financial interpretation of the null model with inactivity periods described in Section \ref{method2} is the following: Brokers typically use for hidden order splitting a VWAP (Volume Weighted Average Price) algorithm \cite{kissel}, which is a simple trading protocol in which the trader splits the order in pieces and trades it incrementally in order to keep approximately constant the ratio, $\alpha_{V}$, between the traded volume by him and the total volume simultaneously traded on that stock in the market. We assume that a VWAP strategy is well represented by a Poisson process with a rate related to the participation rate of the order. The noise in the inventory takes into account the fact that the broker is acting on behalf of many small customers while working the large order. Finally, the inactivity period takes into account the possibility that the broker trades with a bursting activity corresponding to hidden orders and separated by inactivity periods, as recently shown empirically \cite{toth}.

The database we study is the on-book (SETS) market of the London Stock Exchange (LSE), for the period of January 2002 through December 2004. The data set contains all orders placed together with the participant code of the market member placing them. A member can both act as a broker, i.e., handling trades for other institutions that are not members of the market, and may trade for its own account. Thus a single membership code may lump together trades from many different institutions. 
We call the segments that the statistical methods identify, patches. Patches with a well-defined direction (mostly buying or mostly selling) can be understood as hidden orders. In the following we present results from empirical financial data applying both the local time t-test (as in Vaglica et al. \cite{vaglica2008}) and our composite test for segmenting the time series.

We study the inventory variation of a market member in \emph{global transaction time}. This means that for each transaction in which the studied participant is not involved we set his inventory variation to 0, and when he makes a transaction we  put $+v$ for a buy trade and $-v$ for a sell trade. Here $v$ is the volume of the transaction in number of shares. In this way, the time series of the inventory variation contains all information on the participation rate of the agent.

We present here results on the data of AstraZeneca (AZN), a highly liquid stock of the LSE.
We study the 50 most active market members for this period. We consider patches (hidden orders) with a well defined direction. For this reason, as in Refs. \cite{vaglica2008,moro2009}, we apply some filters to the patches found by the methods. Specifically, we define directed patches as those where at least $75\%$ of the volume of transactions inside the patch have the same direction. Furthermore we only study patches that are constituted of at least $10$ transactions. It is also important that in case of financial data, when applying the rate test, we only check for inactivity periods that are at least 50 transactions long. In real time, this corresponds on average to 20 minutes of trading. 

As expected the composite test identifies more directional patches than the local time t-test. In case of AZN the original algorithm determines $3702$ directed patches, while the composite test finds $10613$. This difference suggests that the local time t-test segmentation method tends to stick several patches together, not taking into account changes in the rate and the presence of inactivity periods. This problem is taken into account by our composite test. Thus we expect a better resolution of the patches and less intrinsic error in the detection.

As characterizing statistics of the directed patches we study the number of transactions made inside the patches, $N$, the length of directed patches in global transaction time, $T$, the participation rate in the patch, $\alpha_{V}$,  and the fraction of transactions made via market orders in the patch ($f_{MO}$).
In Table \ref{table1} we summarize some basic statistics of the results, showing the number of directed patches found with each  method and the average of the above measures computed over all directed patches found.

\begin{table}
\centering
\caption{Summary statistics of the local time t-test and the composite test applied to financial data. The rows are number of directed patches found (num),
the average number of transactions made in a directed patch ($\overline N$), the average length of a directed patch in
global transaction time ($\overline{T}$), the average participation rate in directed patches ($\overline \alpha_{V}$),
and the average fraction of market orders in directed patches ($\overline{f}_{MO}$).}
\label{table1}       
\smallskip
\begin{tabular}{lll}
\hline\noalign{\smallskip}
measure & Local time t-test & Composite test\\
\noalign{\smallskip}\hline\noalign{\smallskip\smallskip}
\smallskip\smallskip
num & 3702 & 10613\\
\smallskip\smallskip
$\overline{N}$ & 171.6 & 64.6\\
\smallskip\smallskip
$\overline{T}$ & 1267 & 286\\
\smallskip\smallskip
$\overline{\alpha_{V}}$ & 0.15 & 0.23\\
\smallskip\smallskip
$\overline{f}_{MO}$ & 0.48 & 0.46\\

\noalign{\smallskip}\hline
\end{tabular}
\vspace*{5cm}  
\end{table}

\begin{figure}
\begin{center}
\includegraphics[angle=0,width=0.450\textwidth]{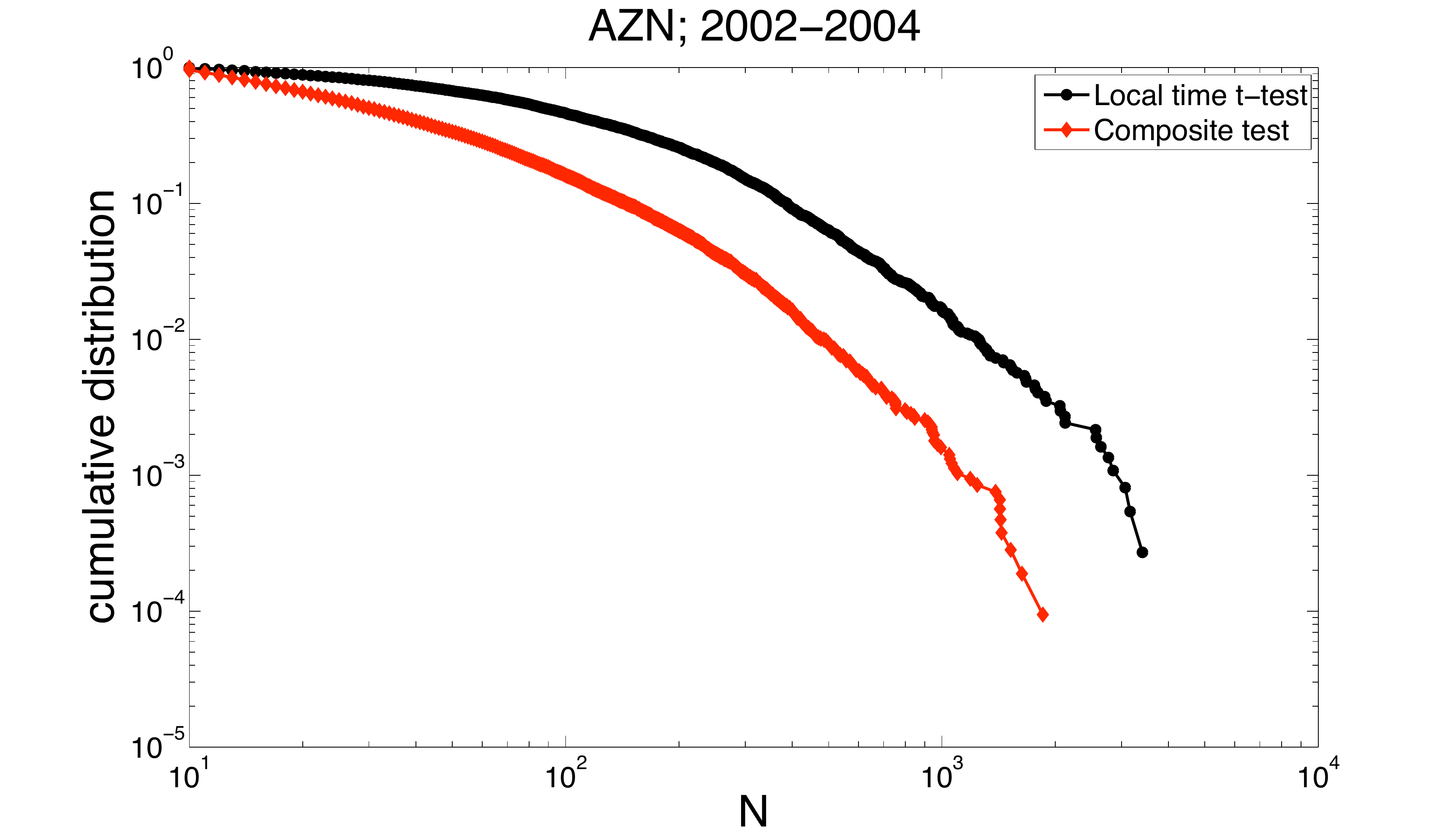}
\caption{\label{fig:N} (Color online) Cumulative distribution of the patch length for directed patches found by the local time t-test (black circles) and
the composite test (red diamonds).}
\end{center}
\end{figure}

Figure \ref{fig:N} shows the cumulative distribution of the number of transactions $N$ in the directed patches. We can see that the distribution
of patch length found using the composite test has a smaller variance than that found with the local time t-test. 
In both cases we find fat tailed distributions.
The average number of transactions constituting a directed patch in case of the local time t-test is $171.6$, while in case of the composite test it is $64.6$.
We find a very large difference in the length of directed patches in global transaction time (see Table \ref{table1}): in case of the local time t-test $\overline{T}=1267$ transactions, which is almost one trading day, while for the composite test $\overline{T}=286$ transactions.

\begin{figure}
\begin{center}
\includegraphics[angle=0,width=0.450\textwidth]{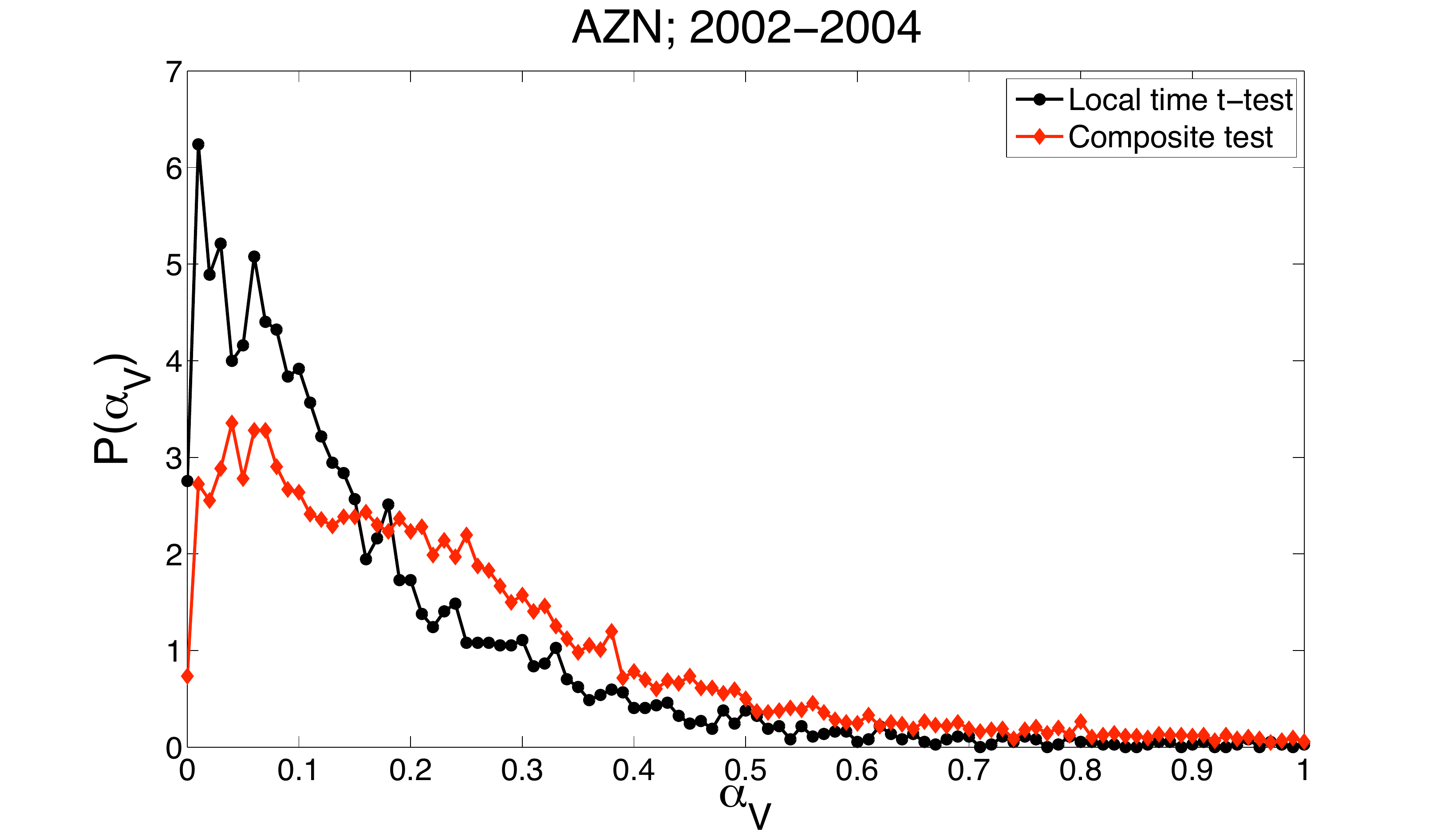}
\caption{\label{fig:alfa} (Color online) Distribution of the participation rate for directed patches found by the local time t-test (black circles) and
the composite test (red diamonds).}
\end{center}
\end{figure}

Figure \ref{fig:alfa} shows the distribution of the participation rate for the directed patches found by the two methods.
The distribution for the directed patches found by the local time t-test is peaked close to zero, with an average of $0.15$, while the distribution of those
found by the composite test has a maximum at a higher value, and decays slower, having an average of $0.23$. 
A simple reason for this difference might be the fact that the composite test finds shorter patches on average and in case of shorter patches it is more likely to find a high participation rate than for long patches (it is very hard to trade with a high participation rate for long periods). Also, if the patches detected with the local time t-test are composed by several shorter patches detected with the composite test and interspersed by inactivity periods, it is clear why the rate is larger in the new than in the local time t-test. 

\begin{figure}
\begin{center}
\includegraphics[angle=0,width=0.450\textwidth]{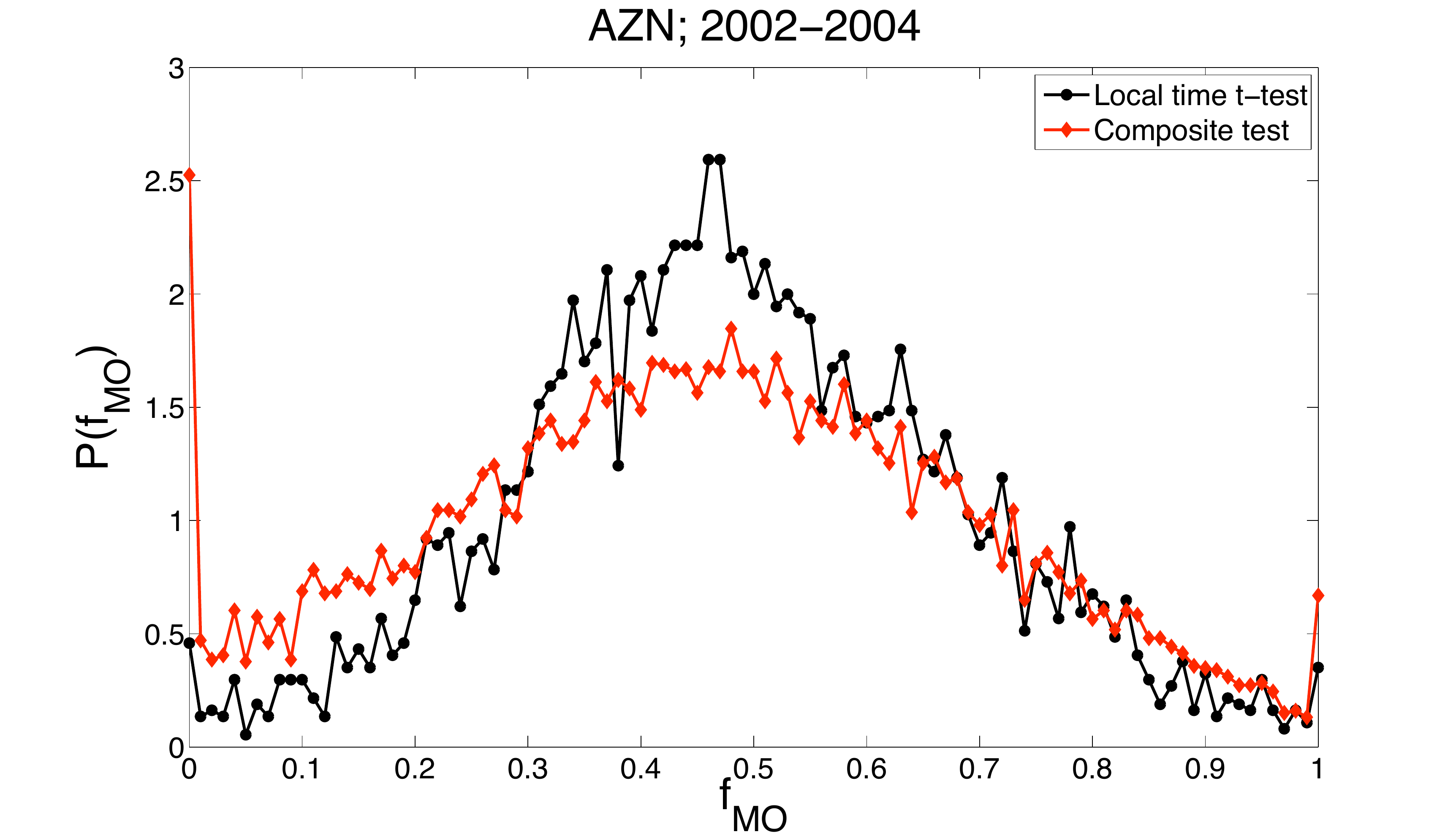}
\caption{\label{fig:fMO} (Color online) The distribution of the ratio of market orders for directed patches found by the local time t-test (black circles) and
the composite test (red diamonds).}
\end{center}
\end{figure}

Figure \ref{fig:fMO} shows the distribution of $f_{MO}$ for the two methods. Both distributions are roughly symmetric 
around $f_{MO}=0.5$. However, we find that in case of the patches found by the local time t-test there is a higher weight
of the distribution in the middle and lower values at the extremes. In case of the composite test the weight in the middle is lower and 
there are large values for the extreme cases, i.e. $f_{MO}=0$ or $f_{MO}=1$. For $f_{MO}=0$ we find the highest value
of the distribution. Again these differences are probably due to the fact that the composite test finds shorter patches. In case of
shorter patches, it is more probable that the participant sticks to one type of order (in our case limit orders).

\subsection{Comparing the segmentations}

An important question is to know the differences between the segmentations found by the two methods. To understand if the composite test
is a ``refinement'' of the local time t-test (finding roughly the same patches and cutting them further into pieces) or if it makes an entirely different 
segmentation, we study the distance of the cuts made by the local time t-test from the nearest cut made by the composite test. We find that
this distance is on average one fourth of the distance that would be found if the two cuts were made independently and on average 
90\% of the distances are lower than what we can expect in the independent case. 
This means that usually there is a cut made by the local time t-test not far from the cuts made by the composite test. 

\begin{figure}
\begin{center}
\includegraphics[angle=0,width=0.450\textwidth]{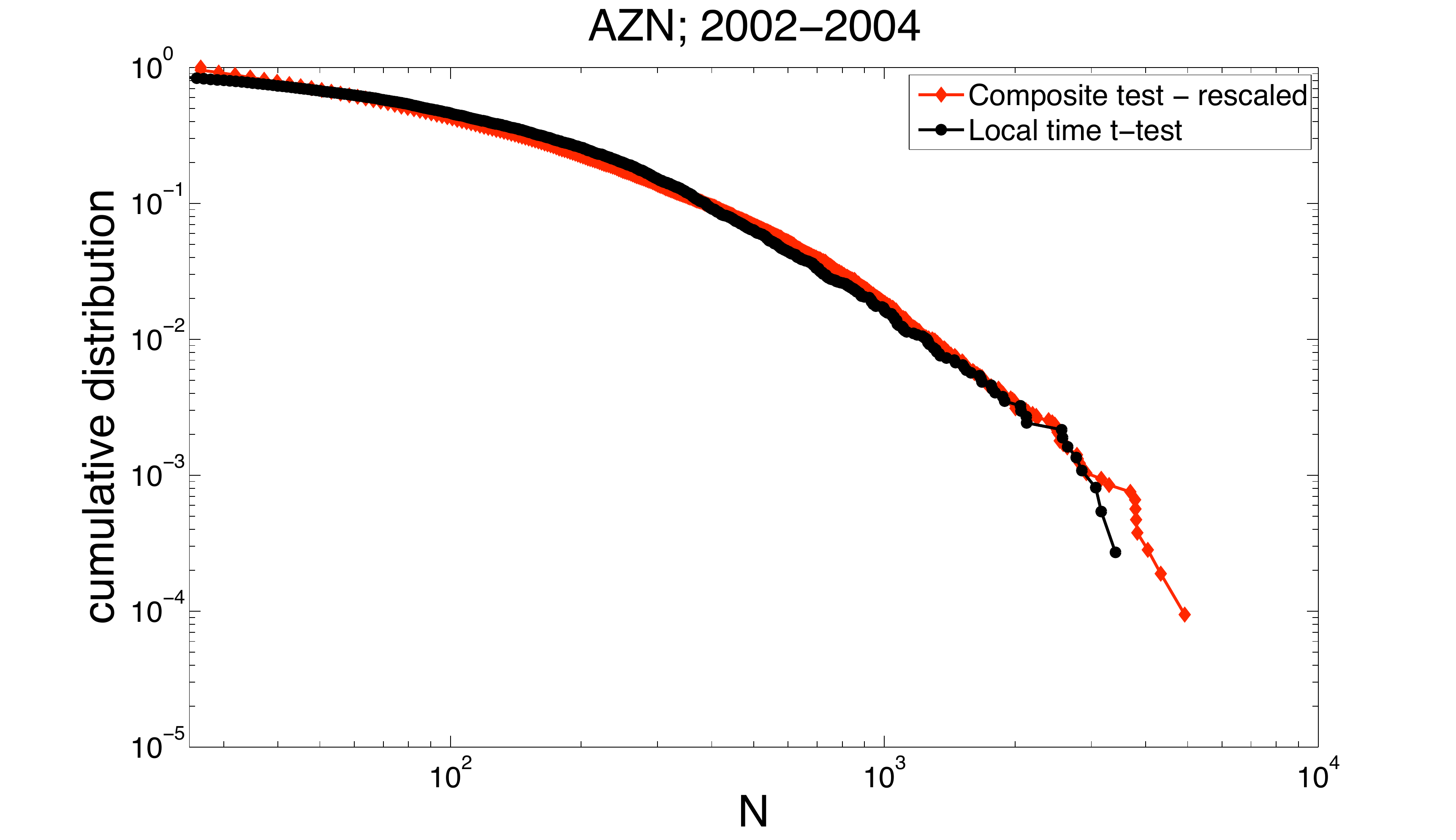}
\caption{\label{fig:N_rescaled} (Color online) Cumulative distribution of the patch lengths for directed patches found by 
the local time t-test (black circles) and
the rescaled cumulative distribution of the patch lengths found by the composite test (red diamonds).}
\end{center}
\end{figure}

We also measured the ``nesting'' of the two methods. Specifically, we measure the fraction of patches found by the composite test that are entirely contained in a patch found by the local time t-test. For this nesting coefficient for AZN we find a value of $0.87\pm 0.09$. This high value suggests that the composite test can partially be seen as a refinement of the local time t-test. In Figure \ref{fig:N_rescaled} we plot the cumulative distribution of N for the two methods (similarly
to Figure \ref{fig:N}), but rescaling the N values of the composite test by the ratio of $\overline{N}$: $171.6/64.6\approx 2.65$ (see Table \ref{table1}).
The two distributions are similar, suggesting again that the composite test is a homogeneous refinement of the local time t-test.

We study the similarity of the two segmentations by computing the Jaccard index of the two patch structures. For the 50 studied brokerage codes we
find a global Jaccard index of $J=0.24\pm 0.21$. This average is interestingly not far from the value that we expect if the composite test were a homogeneous refinement of the local time t-test. In fact the Jaccard index between a partition and its refinement, where each patch in the first partition is divided in $M$ patches of equal length, is $1/M$. In our case this would lead to  $\overline{T}_{composite}/\overline{T}_{local time}=286/1267\approx 0.22$.

\section{Conclusions}
\label{conclusions}

We presented two related methods for segmenting time series that show non-stationarity in both time (rate) and intensity. Specifically our segmentation method has been developed and tested on regime switching models of compound Poisson processes. In this type of process the events occur as in a Poisson process and each event is associated with a signal described by an independent and identically distributed stochastic variable. Moreover, the time series is organized in patches where the parameters of the model are constant within each patch, while they are different in different patches. We have shown that our algorithms perform quite well in segmenting the simulated time series in a wide range of possible parameter distributions.  

Even if the algorithm has been designed and tested for regime switching models of compound Poisson processes, it can be used for more general regime switching models of marked point processes. This is due to the non parametric nature of the segmentation method. Therefore we expect that the method performs quite well in segmenting regime switching time series, where in each patch (regime) the signal is a point process with an intensity associated to each event. 
Examples of time series described by marked point processes  can be taken from a wide range of phenomena, for example, in physiological (neuron spikes), geophysical (earthquakes), astrophysical (solar flares), and socioeconomic (financial transactions) systems.

We also presented an application of our segmentation algorithm to financial  time series. Specifically, we consider the segmentation of the time series describing the inventory of market members. The method looks for periods of time when a market member consistently buys or sells at a roughly constant rate. We postulate that these periods correspond to hidden orders, i.e. large orders that are split and traded incrementally over a long period of time. Hidden orders are an important element for a proper characterization of order flow, which in turn is one of the main ingredients needed to understand the price formation process. 
Given the privacy issues related to the trading activity of market investors, the detection of hidden orders at a financial market scale can be performed only with the aid of statistical methods. Some of the recent efforts in this direction include the methods in Refs. \cite{vaglica2008,moro2009,gerig,hmm}.

The method we introduced in this paper is the first to consider the segmentation in global transaction time, i.e. it is the first that takes into account the trade rate of a given market member. By contrast the other existing methods work in local time, i.e. they discard all the information related to the trading rate.    
When we compare the statistics of the patches found by our algorithm with those found by the algorithm of  Refs. \cite{bernaola-galvan2001,vaglica2008} we find almost three times more patches. This suggests a better resolution in the detection of hidden orders. Comparing the results of the two methods we find evidence that the segmentation of the composite test can be understood as a refinement of the patches found by the local time t-test.

\section*{Acknowledgments} B.T and F.L acknowledge Gabriella Vaglica for useful discussions. F.L. acknowledges financial support from the PRIN project 2007TKLTSR  ``Computational markets design and agent-based models of trading behavior".
J.D.F. and B.T. acknowledge support from NSF grant HSD-0624351. 
Any opinions, findings and conclusions
or recommendations expressed in this material are
those of the authors and do not necessarily reflect the views of
the National Science Foundation.

\end{document}